\documentclass[prd,aps,superscriptaddress, floats,floatfix,eqsecnum,nofootinbib]{revtex4}
\usepackage{amsmath,amssymb,verbatim,graphicx,rotating}
\usepackage{psfrag}
\usepackage{hyperref}

\newcommand{\be}{\begin{equation}}
\newcommand{\ee}{\end{equation}}
\newcommand{\bea}{\begin{eqnarray}}
\newcommand{\eea}{\end{eqnarray}}

\usepackage{graphicx}
\usepackage{dcolumn}
\usepackage{bm}

\begin{document}

\title{Microcanonical model for a gaz of evaporating black holes and strings, scattering amplitudes and mass spectrum }

\author{D. J. Cirilo-Lombardo}
\email{diego@theor.jinr.ru} \affiliation{Bogoliubov Laboratory of Theoretical Physics, JINR Dubna, Moscow region, Joliot-Curie 6, 141980 Russia} \affiliation{Observatoire de Paris, LERMA. Laboratoire
Associ\'e au CNRS UMR 8112.
 \\61, Avenue de l'Observatoire, 75014 Paris, France.}
\author{N. G. Sanchez}
\email{Norma.Sanchez@obspm.fr} \affiliation{Observatoire de Paris,
LERMA. Laboratoire Associ\'e au CNRS UMR 8112.
 \\61, Avenue de l'Observatoire, 75014 Paris, France.}

\date{\today}

\begin{abstract}
We study the system formed by a gaz of black holes and strings within a microcanonical formulation. The density of mass levels grows asymptotically as $\rho (m)\approx (d\ m_{i}+ b m_{i}^{2})^{-a}\ e^{\frac{8\pi}{2}(d\
m_{i}+ b m_{i}^{2})}$, $(i= 1,..., N)$. We derive the microcanonical content of the system: entropy, equation of state, number of components $N$, temperature $T$ and specific heat. The pressure and the specific heat are negative reflecting the gravitational unstability and a non-homogeneous configuration. The asymptotic behaviour of the temperature for large masses {\it emerges} as the Hawking temperature of the system (classical or semiclassical phase) in which the classical black hole behaviour dominates, while for small masses (quantum black hole or string behavior) the temperature becomes the string temperature which emerges as the critical temperature of the system. At low masses, a phase transition takes place showing the passage from the classical (black hole) to quantum (string) behaviour. Within a microcanonical field theory formulation, the propagator describing the string-particle-black hole system is derived and from it the interacting four point scattering amplitude of the system is obtained. For high masses it behaves asymptotically as the degeneracy of states $\rho (m)$ of the system (ie duality or crossing symmetry). The microcanonical propagator and partition function are derived from a (Nambu-Goto) formulation of the N-extended objects and the mass spectrum of the black-hole-string system is obtained: for small masses (quantum behaviour) these yield the usual pure string scattering amplitude and string-particle spectrum $M_n\approx \sqrt{n}$; for growing mass it pass for all the intermediate states up to the pure black hole behaviour. The different black hole behaviours according to the different mass ranges: classical, semiclassical and quantum or string behaviours are present in the model.

\end{abstract}
\maketitle \tableofcontents
 
\maketitle

\section{Introduction and results}

We study the system composed by a gaz of black holes and strings. This 
problem is interesting by its own, it has fundamental and practical physical interest for several reasons. A gaz of primordial black holes and strings could have been formed, existed and decayed in the early universe. In the lack of a microscopic quantum dynamical description of the late stages of black hole evaporation and of a tractable quantum string field theory, the gaz of black holes and strings presents several appealing physical and tractable features. In the early stages of black hole evaporation, most of the emission is in the string massless modes and in the form of Hawking radiation, namely at the Hawking temperature $T_H$, corresponding to the semiclassical evaporation phase [1,2]. As evaporation proceeds, the semiclassical Hawking description breaks down, the quantum string emission is in the most excited massive states, at the string temperature $T_0$, the excited emitted strings undergo a phase transition into a quantum string condensate which then decays as quantum strings do in the usual pure (non mixed) quantum string emission [1,2]. Such phase transition represents the (non perturbative) back reaction effect of the quantum emitted strings on the black hole, (that is when it is not more possible separate the black hole from its emission). The Hawking temperature and the string temperature are the same concept in the two different gravity regimes (semiclassical and quantum regimes) respectively [1-4]. \\

Black holes emit strings, and in their quantum regime, quantum black holes turn to be strings or behave like them, their intrinsic temperature, decay and quantum mass spectrum are similar; on the other hand, strings can, under certain conditions of high excitation and unstability, collapse into black hole states. Thus, a system composed by both, black holes and strings, allow to describe the different phases and black hole behaviours: classical or semiclassical black holes, ie black holes at their early phases of evaporation, emission of strings by the black holes, and quantum black holes at the last stages of their evaporation.\\

In this paper we study the gaz of black holes and strings within a microcanonical description; the microcanonical approach is the most physical compelling for this system.   We first study it within an ideal gaz statistical mechanical ensemble; we also discuss the N-body interacting amplitudes by analogy with the string and dual models. We then study the system within a Quantum Field Theory microcanonical formulation (with interactions) for the propagator, scattering amplitude and partition function. Finally, we derive the microcanonical propagator of the system from the Nambu-Goto action for N- extended objects, which takes into account the non-local effects.\\

Previously, the statistical properties of the gaz of strings alone [8]-[11] or the gaz of black holes alone [12]-[14] were considered, but the system composed of both extended objects black holes and strings which allow to cover a rich class of situations was 
never considered before.\\

We compute the microcanonical partition function $\Omega _{n}(n,E,V)$ for the gaz of black holes and strings with the density of mass levels having the following asymptotic behaviour 
\begin{equation}
\rho (m)\approx \ (d\ m_{i}+  b\ m_{i}^{2})^{-a}\ e^{\frac{8\pi}{2}(d\
m_{i}+ \ b m_{i}^{2})}  \label{eq:1}
\end{equation}
$a$, $b$ and $d$ being constants fixed by the boostrap condition. It appears that the global behaviour of the system depends on two cases: $a=5/2$ and $a\neq 5/2$, but there are relevant generic features common to the both cases and we will stress them here. We find for the microcanonical partition function
\begin{equation}
\Omega _{n}(n,E,V) = e^{\frac{8\pi }{2}(d\ E+E^{2}\ b)}e^{g\left(n,E,V\right)} 
\label{eq:1.2}
\end{equation}

where, by using the Stirling approximation, we find for the density of states $g\left(n,E,V\right)$:
\begin{equation}
g\left(n,E,V\right) \simeq -n\ln n + n +n \ln \left[ C \right] + n \ln \left[ H(E)\right]
\end{equation}

$C$ is a constant which depends on different combinations of the parameters, $a$, $b$, $d$ and on $m_0$, the minimal mass in the system $(m_0 << E)$, and $H(E)$ is a function which we compute explicitely in the two cases of interest: $a\neq 5/2$ and $a=5/2$. In this last case $H(E)= n/E$.
The first factor in eq.(1.2) is the string-black hole
microcanonical partition function and the other factors are deviations from it. The leading part of $\Omega _{n}(n,E,V)$ is corrected by the $n$ dependent terms. \\

We study the microcanonical content of the system. The entropy is read from eqs (1.2),(1.3):
\begin{equation}
S \simeq \frac{8\pi}{2}(Ed+E^{2}b) + g\left(n,E,V\right), 
\end{equation}
and we find for the Temperature: 
\[
T= \frac{T_0}{1 + 8 \pi E bT_0 + F(E)T_0}
\]
with $F(E)/E \rightarrow 0$ for $E\rightarrow \infty$ and $F(E) \rightarrow 0$ for $E\rightarrow 0$. The function $F(E)$ is explicitely obtained in the two cases $a=5/2$ and $a\neq 5/2$. The asymptotic large energy regime and the small energy regime of the temperature {\it emerge} respectively as the Hawking temperature $T_H$ and the string temperature $T_0$ of the system : 
\[
 T(E\rightarrow \infty) = \frac{1}{8 \pi E b}= T_H  ~~~~~,~~~~~~T(E\rightarrow 0)= T_0 ,~~~~   (T_0=1/4\pi d)
\]

The pressure is negative, as a function of $N$:
\[
\left. PV\right| _{a\neq 5/2}= -\frac{N\ T}{\left( 5/2-a\right) }\ \ln N 
\]
while it is positive for $a=5/2$. For $a\neq 5/2$ the specific heat $C_{v}$ is negative, for low energies it behaves as ~$ C_{v} \approx \frac{-1}{8\pi b T_{0} ^{2}}%
\approx const$,~while for high energies $C_{v}$ vanishes, showing up the asymptotic stability point of the system. The regimes in which the pressure becomes negative reflect the presence of gravothermal unstable behaviour, which is accompassed by the behaviour of the specific heat. The changes in the local energy density given by the
microcanonical ensemble allow the negative value for the specific heat $C_{v}$.
The canonical ensemble (which takes the average of the energy) does not take
care of the local fluctuations of the energy density and, for instance, both
approaches (canonical and microcanonical) are not equivalent in this case. The behaviour of the microcanonical physical magnitudes indicate that for small energy a phase transition takes place showing the passage from macroscopic (classical black hole) to microscospic (string) behaviour;\\
 
We also check for our statistical model the duality (crossing symmetry) of the scattering amplitudes characteristic of string and dual systems: We evaluate the two-body (and $n$-body) amplitude $N_n$ of the string-black hole system and find that $N_n \approx \rho (m)\approx (d\ m_{i}+ b m_{i}^{2})^{-a}\ e^{\frac{8\pi}{2}(d\
m_{i}+ b m_{i}^{2})}$, $(i= 1,..., N)$ for $ m\rightarrow \infty $. That is, the number of open channels does grow in paralell with the degeneracy of states, the number of resonnances, as energy is increased. \\

In order to go further in the understanding of this system, we study it within a microcanonical field theory (or thermo field) formulation. We compute (in momentum space) the physical component of the microcanonical propagator $\Delta ^{11} _E (k)$ and from it the four point scattering amplitude $A$ of the system, which expresses as the sum of two parts: $A = A_{str-particle}+ A_{str-bh}~$, the first term corresponds to the usual string-particle spectrum (Veneziano amplitude), the second one describes the {\bf new} string-black hole component of the amplitude, which we compute explicitely, eq(5.4). For high mass, $A_{str-bh}\cong \rho \left(m\right) $ and we recover that it increases exponentially as the density of states for the string-black-hole system. We then derive the microcanonical partition function $\Omega (E,V)$ from the computed field propagators and transition  amplitude $A$, relating  $\ln \Omega (E,V)$ to the trace of the connected part of the imaginary part of the full propagator. 

\bigskip
Finally, we give a Nambu-Goto formulation of the N-extended body system composed by black holes and strings and derive from it the microcanonical propagator 
\begin{equation}
D_{E}(k,m) =\frac{\delta \left(E\right)}{\omega^{2}-k^{2}-m^{2}+i\varepsilon }- D_{E~(str-bh)}(k,m)
\label{eq:1.4}
\end{equation}

\begin{equation}
D_{E~(str-bh)}(k,m)= 8\pi i\alpha \delta \left( \omega ^{2}-k^{2}-m^{2}\right) \sum_{l=1}^{\infty } \frac{K_{-1}\left( \alpha \left|
l\omega _{k}-E\right| \right) }{E^{2}}\frac{\Omega \left( E-l\omega
_{k}\right)}{\Omega \left( E\right)}\theta \left(E-l\omega _{k}\right) 
\end{equation}

where $l,\omega _{k}$ and $\theta\left( E-l\omega _{k}\right)$ are the mode number, the dispersion relation and the step function respectively, and $K_{-1}$ is the Mac Donald's function. The first term in the microcanonical propagator eq.(1.6) is the usual Feynman
propagator, the second one is the new microcanonical part $D_{E~(str-bh)}(k,m)$ describing the string-black hole component. From this term, we obtain for $E \rightarrow 0$, being $E = M$ the total mass of the system, the pure string amplitude (Veneziano amplitude). When $E$ grows it pass for all intermediate states up to the pure black hole behaviour.  This propagator includes all non-local effects from the $N$-bodies of the system considered as extended objects and from its derivation as in the string Nambu-Goto formulation. For instance, the propagator eq.(1.6) becomes the (point particle) Quantum Field Theory propagator  $\Delta ^{11} _E (k)$ when the string becames point-particles and the Hamiltonian is quadratic in the momenta (constrained particle Hamiltonian). We analyze the spectrum (singular points of the microcanonical part). In the excited regime $l\rightarrow M/\omega _{k}$,  the spectrum becomes 
\begin{equation}
\left( \frac{8\pi d}{a}\right)^{2} M ^2 \simeq n\ln (\frac{8\pi}{a}) -\ln 2~~~,~~~~ n>>1
\label{eq:1.5}
\end{equation}
which is like the usual quantum string spectrum, that is the mass of quantum black holes is quantized in the same way as quantum strings.\\

This study allowed us to derive a wide class of properties and physical magnitudes of the system composed by black and strings, covering the different mass ranges and the classical, semiclassical and quantum behaviours of the system.

\bigskip

This paper is organized as follows: In section II we formulate the statistical mechanics of the gaz of strings and black holes in the microcanonical ensemble and derive the partition function of the system; in section III we compute from it the physical magnitudes of the system: entropy, temperature, pressure and specific heat, and analyze their properties. In section IV we discuss the duality (crossing) property of the scattering amplitudes of the system in analogy with that of string and dual models. In section V we provide a quantum field theory description of the system (propagator and scattering amplitudes within the operator approach). In section VI we describe the system of N-black holes and strings as N-extended objects in the Nambu-Goto formulation, and derive in this context the microcanonical propagator and mass spectrum.


\section{Statement of the problem}

In the microcanonical ensemble, we consider the following density of states
for a configuration with $n$ black holes and strings 
\begin{equation}
\Omega _{n}(n,E,V)=\sum_{n=0}^{\infty }\left[ \frac{V%
}{\left( 2\pi \right) ^{3}}\right] ^{n}\frac{1}{n!}\prod_{i=1}^{n}%
\int_{m_{0}}^{\infty }dm_{i}\int_{-\infty }^{\infty }\rho (m)dp_{i}^{3}\ \
\delta \left( E-\sum_{i=1}^{n}E_{i}\right) \ \delta ^{3}\left(
\sum_{i=1}^{p_{i}}\overline{p}_{i}\right)  \label{eq:1}
\end{equation}
where we make the following considerations :
\\

(i) the density of mass levels has the following asymptotic behaviour 
\begin{equation}
\rho (m)\approx c\ (d\ m_{i}+m_{i}^{2}\ b)^{-a}\ e^{\frac{8\pi }{2}(d\
m_{i}+m_{i}^{2}\ b)}  \label{eq:2}
\end{equation}
$a,$ $b,$ $c$ and $d~$being positive parameters whose meaning will be clear
in the sequel (they are fixed by the boostrap condition eq. (2.21)). This
density of mass levels takes into account both the string component and the
black hole component of the system.

(ii) a particle-like relation of dispersion $E_{i}=\sqrt{m_{i}^{2}+\left| 
\overline{p}_{i}\right| ^{2}}\ \ \ \ \ \ \ \ \ (G=c=1)$

$E_{i}$ being the energy of the i$^{th}$ black hole with linear momentum $%
\overline{p}_{i}.$

(iii) the conditions: $b^{-1}<m_{i}^{2}$ and $d^{-1}<m_{i}$ , which are the usual ones in statistical hadronic systems. Here these conditions mean that the
constants $b$ and $d$ arise naturally from the statistical system considered
as delimiters of the kinetic status of the particles.

(iv) the mass $m_{0}$ is the mass of the less massive component (black
hole or string) in the gas, namely $m_{0} << M$, being $M$ the total mass of the system (black holes and strings).
\\

(v) the factor $\delta ^{3}\left( \sum^{p_{i}}_{i=1} \overline{p}_{i}\right) $
can be neglected (over-all momentum conservation).

\bigskip Therefore, the expression (2.1) takes the following form 
\begin{equation}
\Omega _{n}(n,E,V)=\left[ \frac{V}{\left( 2\pi \right) ^{3}}\right] ^{n}%
\frac{1}{n!}\prod_{i=1}^{n}\int_{m_{0}}^{\infty }\ dm_{i}\int_{-\infty
}^{\infty }\rho (m)dp_{i}^{3}\ \ \delta \left( E-\sum_{i=1}^{n}E_{i}\right)
\   \label{eq:3}
\end{equation}

We rewrites the energy $E_{i}=m_{i}+Q_{i}\ $in terms of the kinetic energy $%
Q_{i}$ and consider the dominant part of it in the usual form $\left(
Q_{i}\simeq \frac{p_{i}^{2}}{2m_{i}}\right) $. Then,\newline
\begin{equation}
\prod_{i=1}^{n}e^{\frac{8\pi }{2}(d\ m_{i}+m_{i}^{2}\ b)}=e^{\frac{8\pi }{2}%
(d\ E+E^{2}\ b)}\prod_{i=1}^{n}e^{-8\pi b(\frac{Q_{i}^{2}}{2}+m_{i}Q_{i}\
)}e^{-\frac{8\pi d}{2}Q_{i}}  \label{eq:4}
\end{equation}

and define 
\[
I_{i}\left( \Lambda _{i}\right) \equiv \int_{m_{0}}^{\Lambda _{i}}c\ (d\
m_{i}+m_{i}^{2}\ b)^{-a}\ dm_{i}\int e^{-8\pi bQ_{i}\left[ \frac{Q_{i}}{2}%
+m_{i}\left( 1+\frac{d}{2bm_{i}}\right) \right] }\ dp_{i}^{3}\ \ 
\]
\newline
where the \ cut-off $\Lambda _{i}$ on momentum integration for $Q_{i}^{2}\approx
b^{-1}\approx d^{-2}$ is naturally introduced by the condition (iii). We
then have 
\begin{eqnarray*}
I_{i}\left( \Lambda _{i}\right) &\equiv &\int_{m_{0}}^{\Lambda _{i}}c\ (d\
m_{i}+m_{i}^{2}\ b)^{-a}\ dm_{i}\int e^{-\frac{8\pi b}{2}p_{i}^{2}\left[
\left( \frac{p_{i}}{2m_{i}}\right) ^{2}+\left( 1+\frac{d}{2bm_{i}}\right)
\right] }\ dp_{i}^{3}\ \  \\
&=&\int_{m_{0}}^{\Lambda _{i}}c\ (d\ m_{i}+m_{i}^{2}\ b)^{-a}\ \frac{\sqrt{\pi}}{4}%
\left( \frac{2\pi b}{m_{i}^{2}}\right) ^{-3/4\ }\ 
e^{\ 2\pi b\widehat{m}_{i}^{2}}D_{-3/2}\left( 2\sqrt{2\pi b}\widehat{m}%
_{i}^{2}\right) 
\end{eqnarray*}

where $D_{-3/2}\left( x\right) $ is the Whittaker function. From eqs. (2.2)
and (2.3) we get 
\begin{equation}
\Omega _{n}(n,E,V) =\frac{e^{\frac{8\pi }{2}(d\ E+E^{2}\ b)}}{n!}\left[ 
\frac{V}{\left( 2\pi \right) ^{3}}c\left( \frac{\pi }{b^{3}}\right) ^{\frac{1
}{4}}\right] ^{n}\prod_{i=1}^{n}\int_{m_{0}}^{\infty }\left( \frac{\widehat{m}_{i}}{2
}\right) ^{3/2}(d\ m_{i}+m_{i}^{2}\ b)^{-a}\ \Upsilon \ \delta (
\sum_{i=1}^{n}m_{i}-E)~ dm_{i}  
\label{eq:5}
\end{equation}
where 
\begin{equation}
\widehat{m_{i}}=m_{i}+\frac{d}{2b}  \label{eq:6}
\end{equation}
and
\begin{equation}
\Upsilon \equiv \Gamma \left( -1/4\right) \ _{1}F_{1}\left[ 3/4,1/2,4\pi b 
\widehat{m}_{i}^{2}\right] +2\sqrt{2\pi b}\widehat{m_{i}}\ \Gamma \left(
1/4\right) \ _{1}F_{1}\left[ 5/4,3/2,4\pi b \widehat{m}_{i}^{2}\right] 
\label{eq:7}
\end{equation}
$_{1}F_{1}$ being the confluent hypergeometric function. Let us now define
\begin{equation}
\prod {}^{n}\left(E\right) \equiv \prod_{i=1}^{n}I_{i}\left(\Lambda
_{i}\right) =\frac{e^{\frac{8\pi }{2}(d\ E+E^{2}\ b)}}{n!}\left[ \frac{c}{
2^{3/2}}\left( \frac{\pi }{b^{3}}\right) ^{\frac{1}{4}}\right] ^{n}
 \int_{nm_{0}}^{E}dX\prod_{i=1}^{n}\int_{m_{0}}^{\infty}dm_{i}\left( 
\frac{\widehat{m}_{i}}{2}\right)^{3/2}(d\ m_{i}+m_{i}^{2}\ b)^{-a}\
\Upsilon \ \delta ( \sum_{i=1}^{n}m_{i}-X)  
 \label{eq:8}
\end{equation}
such that 
\begin{eqnarray}
\Omega _{n}(n,E,V) =\frac{e^{\frac{8\pi }{2}(d\ E+E^{2}\ b)}}{n!}\left[ 
\frac{V}{\left( 2\pi \right) ^{3}}\right] ^{n}\frac{d\left( \prod
{}^{n}\left( E\right) \right) }{dE}   \equiv &e^{\frac{8\pi }{2}(d\ E+E^{2}\ b)}e^{g\left( n,E,V\right) } 
\label{eq:9}
\end{eqnarray}
where the constraint is obviously 
\begin{equation}
\sum \Lambda _{i}=E  \label{eq:10}
\end{equation}

The maximum contribution to $\prod {}^{n}\left( E\right) $ is obtained when
all the $\Lambda _{i}$\ are of the order $\frac{E}{n}$. We have $I\left(
\Lambda _{i}\right) $ with the hypergeometric confluent functions in
assymptotic form for $2\sqrt{2\pi b}\widehat{m_{i}}>>1$ since the condition
(iii). We get in this manner 
\begin{equation}
\prod {}^{n}\left( E\right) =\left[ \frac{c}{2^{3/2}}\left( \frac{\pi }{b^{3}
}\right) ^{\frac{1}{4}}\right] ^{n}
 \prod_{i=1}^{n}\int_{m_{0}}^{\infty }( 1+\frac{d}{2bm_{i}}
) ^{-3/2}\ (d\ m_{i}+ m_{i}^{2}\ b)^{-a}\left[1-\frac{\frac{3
}{2}\frac{5}{2}}{2.4.(2\pi b)\widehat{m}_{i}^{2}}+...\right] \delta
(\sum_{i=1}^{n}m_{i}-E)\ dm_{i}   
\label{eq:11}
\end{equation}
From the above expression, two different behaviours show up depending on the value of $a$:

(1) $a\neq 5/2:$
\begin {equation}
\prod {}^{n}\left( \Lambda _{i}\right) =\left[ \frac{c}{2^{3/2}}\left( \frac{%
\pi }{b^{3}}\right) ^{\frac{1}{4}}\right] ^{n}\left\{ \left( \frac{2b}{d}%
\right) ^{3/2}\frac{1}{5/2-a}\left[ \Xi \left( \Lambda _{i}\right) +\Xi
\left( m_{0}\right) \right] \right\} ^{n}  \label{eq:12}
\end{equation}
where 
\[
\Xi \left( x\right) \equiv x^{3/2-a}\left[ F_{1}\left( 5/2-a,3/2+a,7/2-a,-%
\frac{2bx}{d}\right) \ -\frac{15}{16\pi }\frac{b}{d^{2}}F_{1}\left(
5/2-a,7/2+a,7/2-a,-\frac{2bx}{d}\right) \right]    
\]

(2) $a=5/2:$%
\begin{eqnarray}
\prod {}^{n}\left( \Lambda _{i}\right) &=&\left[ \frac{c}{2^{3/2}}\left( 
\frac{\pi }{b^{3}}\right) ^{\frac{1}{4}}\right]^{n}\left\{ A\left[ arcth%
\sqrt{h\left( \Lambda _{i}\right)}- arcth\sqrt{h\left( m_{0}\right)}\right]
\right. +  \\
&&+\left. \left( \frac{2b}{d}\right) ^{3/2}\left[\frac{1}{\sqrt{h\left(
\Lambda _{i}\right) }}-\frac{1}{\sqrt{h\left( m_{0}\right)}}\right] -\frac{%
15}{32\pi b}\left(2b\right) ^{7/2}\left[ p\left( \Lambda _{i}\right)
-p\left( m_{0}\right) \right]\right\}^{n}  
\end{eqnarray}
with 
\[
h(x) \equiv 1+\frac{2bx}{d}\ \ ;\ \ \ \ A\equiv \frac{15}{32\pi b%
}\left( \frac{2b}{d}\right)^{3/2}\left[\left(\frac{2b}{d}\right)^{2}-1 \right]\ \ ;\ \ \ \ p\left( x\right) \equiv \frac{1}{d^{7/2}h\left( x\right)^{5/2}}\left( \frac{%
1}{5}+\frac{h\left( x\right) }{3}+h\left( x\right)^{2}\right) 
\]

Then, from eqs (2.11) and (2.12)-(2.14), the microcanonical density $g\left(n,E,V\right)$ in the two cases ($a\neq 5/2$ and $a=5/2$ respectively) is given by :
\begin{equation}
e^{g\left(n,E,V\right)}(a\neq 5/2) \approx \frac{1}{n!}
\left[\gamma \left(\frac{2b}{d}\right)^{3/2}\frac{1}{5/2-a}\right]^{n}n \left\{\eta \left[\left(\frac{E}{n}\right)^{3/2-a}-m_{0}^{3/2-a}\right]\right\}^{n-1} \times \nonumber
\end{equation}
\begin{equation}
\times \left[\eta \frac{3/2-a}{E}\left(\frac{E}{n}\right)^{3/2-a}-\frac{5/2-a}{7/2-a}\left(\frac{E}{n}\right)^{5/2-a}\frac{b}{Ed}\left(3-\frac{7b}{d^{2}}\frac{15}{16\pi}\right)\right]  
\label{eq:14}
\end{equation}
\begin{equation}
e^{g(n,E,V)}(a=5/2) \approx \frac{\left[\gamma
q( 1-\frac{15q^{2}}{32\pi b})\right]^{n}}{n!}\left[1+\frac{%
f(m_{o})}{q\left(1-\frac{15q^{2}}{32\pi b}\right)}\right]
^{n-1}\left(\frac{n}{E}\right)  \label{eq:15}
\end{equation}
where 
\begin{eqnarray*}
f(m_{0}) &\equiv &\lambda~arcth\left(\sqrt{h(m_{0})}\right)
+ \left(\frac{2b}{d}\right)^{3/2}\left(1-\frac{1}{\sqrt{h(
m_{0})}}\right)-(2b)^{7/2}\frac{15}{32\pi b}\left[\frac{23}{15 d ^{7/2}}-p( m_{0})\right]
\end{eqnarray*}
\begin{equation}
\gamma \equiv \left[\frac{V}{(2\pi)^{3}}c\left(\frac{\pi}{%
b^{3}}\right)^{\frac{1}{4}}\right] \ \ \ ;\ \ \ \eta \equiv \left( 1-\frac{15}{%
16\pi}\frac{b}{d^{2}}\right) \ \;\ \ \ q\equiv \left(\frac{2b}{d}\right)^{3/2}
\label{eq:16}
\end{equation}
Using the Stirling approximation, the density of states $g\left(
n,E,V\right)$ eqs.(2.15) and (2.16) can be expressed as 
\begin{equation}
g\left(n,E,V\right) _{a\neq 5/2} \simeq -n\ln n+n+n\ln \left[\gamma \left(\frac{2b}{d}\right)^{3/2}\frac{1}{5/2-a}\right] +n \ln \left\{\eta \left[(\frac{E}{n}) ^{3/2-a}-m_{0}^{3/2-a}\right]\right\} +
\end{equation}
\begin{equation}
+ \ln \left[\eta \frac{3/2-a}{E}\left(\frac{E}{n}\right)^{3/2-a}-\frac{
5/2-a}{7/2-a} \left(\frac{E}{n}\right)^{5/2-a}\frac{b}{Ed}\left(3-\frac{7b
}{d^{2}}\frac{15}{16\pi}\right)\right]  
\label{eq:17} 
\end{equation}
and 
\begin{equation}
g\left( n,E,V\right) _{a=5/2} \simeq -n\ln n+n+n\ln \left[\gamma q\left(
1-\frac{15q^{2}}{32\pi b}\right)\right] + (n-1)\ln \left [1+\frac{f m_{o})}{q\left( 1-\frac{15q^{2}}{32\pi b}\right)}\right] + \ln \left(\frac{n}{E}\right)  
\label{eq:18}
\end{equation}
It is interesting to see from eq.(2.20) and eq.(2.16) that when $m_{o}\approx 0$,
then $f\left( m_{o}\right) \approx 1$, and we have 
\[
g\left( n,E,V\right) _{a=5/2}\simeq -n\ln n+n+n\ln \left[ \gamma q \left(1-%
\frac{15q^{2}}{32\pi b}\right) +\gamma \right] +\ln \left(\frac{n}{E}\right) 
\]
which express in a more clear form how the leading part of the microcanonical
density of states $\Omega _{n}(n,E,V)$ is corrected by the $n$ dependent terms 
\[
\Omega _{n}(n,E,V)\simeq e^{\frac{8\pi }{2}(d\ E+E^{2}\ b)}\times \frac{%
ne^{n}}{E}\left( \frac{const.}{n}\right) ^{n} 
\]
It is easy to see that the first factor is the string-black hole
microcanonical partition function and the other factors are deviations from it.
\\

The extent of the parameters $(a,b,c,d)$ depends on the boostrap constraint. If we use the strong condition, explicitly 
\begin{equation}
\lim_{E\rightarrow \infty }\ \left[ \frac{\Omega (E,V)}{\rho (E)}\right]
\rightarrow 1 ~~, \label{eq:19}
\end{equation}
this condition holds for 
\begin{equation}
b=1=d^2 ~~~~, ~~~~c \neq 0 ~~~~ \text{arbitrary},
\label {eq:3.2}
\end{equation}

and allowing two cases for the $a$ parameter: (i) $a\neq 5/2$ with $2\leq a<5/2$ or $a>5/2$; and (ii) $a=5/2$. \\

With the units restoired $ b= t_{Pl}^{-2},d = t_s ^{-1},$, $t_{Pl}$ and $t_s$ being the fundamental Planck temperature and fundamental string temperature respectively: $ t_{Pl}= \sqrt{\hbar c/G }$, $t_s = \sqrt{\hbar c/\alpha'}$.
\bigskip

Let us now discuss the physical microcanonical content of the system.

\section{Microcanonical content of the model}

\subsection{Entropy and Number of components of the system}

In the microcanonical formulation, we have for the entropy of our system
\begin{equation}
S = \ln \Omega _{n}(E,V) = \frac{8\pi }{2}(Ed + E^{2}b)+ g\left( n,E,V\right)
\label {eq.18a }
\end{equation}
From eqs.(2.18)-(2.20) we have 
\begin{equation}
S\left(a\neq 5/2\right) \simeq \frac{8\pi}{2}(Ed+E^{2}b)-n \ln n + n +
n\ln \left[\gamma \left( \frac{2b}{d}\right) ^{3/2}\frac{1}{5/2-a}\right] + n\ln \left\{\eta \left[\left(\frac{E}{n}\right)^{\frac{3}{2}-a}- m_{0}^{3/2-a}\right]\right\}
\end{equation}
\begin{equation}
S\left(a=5/2\right) \simeq \frac{8\pi}{2}(Ed+E^{2}b)-n\ln n + \ln\left(\frac{n}{E}\right) 
+ n\ln \left[\gamma q\left( 1-\frac{15q^{2}}{32\pi b}\right)\right]
+(n-1)\ln \left [1+\frac{f\left( m_{o}\right)}{q\left(1-\frac{15q^{2}}{
32\pi b}\right)}\right]
\end{equation}

where $f\left( m_{o}\right) $ is given by eq.(2.17), and we can see that the
first and second terms in the expressions above correspond to the string
entropy and to the black-hole entropy respectively. The other $n$-dependent and $E$-dependent terms are corrections to the entropy of the system.
\\
The most probable number of components $N$ for the system is obtained by
maximizing the partition function with respect to $n.$ For $a\neq 5/2$ the
result is 
\begin{equation}
n=N\approx \left[ \frac{\widehat{\gamma }\eta (1-\Delta^{^{3/2-a}}) E^{3/2 - a}}{( 5/2 -a -\Delta^{^{3/2-a}})
m_{0}^{-a}}\right]^{\frac{1}{5/2-a}}  \label{eq:20}
\end{equation}
where 
\[
\widehat{\gamma }\equiv \gamma \left( \frac{2b}{d}\right) ^{3/2}\frac{1}{%
5/2-a}\ \ \ ;\ \ \ \ \ \ \Delta \equiv \left( \frac{N\ m_{0}}{E}\right) 
\]
The units of the constant $c$ are such that $\gamma $ is 
dimensionless. The particle number density $N/V$ is a function 
of the energy density $E/V$ only. The energy as a function of $N$ is 
\[
E= \left[\frac{N^{5/2-a} ( 5/2-a-\Delta^{^{3/2-a}} ) }{\widehat{
\gamma }\eta ( 1-\Delta ^{^{3/2-a}}) m_{0}^{a}}\right]^{\frac{1}{3/2-a}} 
\]

We see that $N$ goes to zero for $E\rightarrow Nm_{0}$ (the lowest energy of
the system) with $a<5/2$ or $a\rightarrow 3/2$, while

$N \rightarrow  \infty $ when $\Delta \rightarrow \left( 5/2-a\right) ^{%
\frac{1}{3/2-a}}$~~and ~$a<5/2$.
\\
For $a\neq 5/2$ , from eqs. (2.18) and (3.4), $g\left( N,E,V\right)$ is
expressed as:
\begin{equation}
g\left( N,E,V\right)_{a\neq 5/2} \approx  \left[\frac{\widehat{\gamma }\eta (1-\Delta ^{^{3/2-a}} ) E^{3/2-a}}{(5/2-a-\Delta^{^{3/2-a}}) m_{0}^{-a}} \right]^{\frac{1}{5/2-a}} \left\{1+\frac{1}{5/2-a}\ln \left[\frac{( 5/2+a-\Delta^{^{3/2-a}}) m_{0}^{-a}}{
\widehat{\gamma}\eta(1-\Delta^{^{3/2-a}}) E^{3/2-a}}\right]\right\}  
\label{eq:21}
\end{equation}

which has the following behaviours :
\[
\begin{tabular}{|l|l|l|}
\hline
& $g\left( N,E,V\right) _{E\rightarrow 0}$ & $g\left( N,E,V\right)
_{E\rightarrow \infty }$ \\ 
&  &  \\ \hline
$a<3/2:$ & $ ~~\left[ \frac{\widehat{\gamma }\eta }{m_{0}^{-a}}\right] ^{%
\frac{1}{5/2-a}}E^{\frac{3/2-a}{5/2-a}}\rightarrow 0$ & $ \left[ 
\frac{\widehat{\gamma }\eta }{m_{0}^{-a}}\right] ^{\frac{1}{5/2-a}}E^{\frac{%
3/2-a}{5/2-a}}\rightarrow \infty $ \\ 
&  &  \\ \hline
&  &  \\ 
$a>5/2:$ & $ ~~\left[ \frac{\widehat{\gamma }\eta }{\left( 5/2-a\right)
m_{0}^{-a}}\right] ^{\frac{1}{5/2-a}}E^{\frac{3/2-a}{5/2-a}}\rightarrow 0$ & 
$~~ \left[ \frac{\widehat{\gamma }\eta }{m_{0}^{-a}}\right] ^{\frac{1}{%
5/2-a}}E^{\frac{3/2-a}{5/2-a}}\rightarrow \infty $ \\ 
&  &  \\ \hline
&  &  \\ 
$3/2<a<5/2:$ & $ ~~\left[ \frac{\widehat{\gamma }\eta }{\left(
5/2-a\right) m_{0}^{-a}}\right] ^{\frac{1}{5/2-a}}E^{\frac{3/2-a}{5/2-a}%
}\rightarrow \infty $ & $\left[ \frac{\widehat{\gamma }\eta }{%
m_{0}^{-a}}\right] ^{\frac{1}{5/2-a}}E^{\frac{3/2-a}{5/2-a}}\rightarrow 0$
 
\\ \hline
\end{tabular}
\]
For $a=3/2$, $g\left( N,E,V\right)$ does not depends on
the energy $E$ : 
\begin{equation}
g\left(N,E,V\right)\approx \left(\frac{\widehat{\gamma }\eta }{m_{0}^{-3/2}}\right) \left[1+\ln \left(\frac{m_{0}^{-3/2}}{\widehat{\gamma}\eta}\right)\right]  
\label{eq:22}
\end{equation}
. 

For $a=5/2$, the most probable number of components of the system is 
\begin{equation}
n = N \approx \gamma \left[ f( m_{0}) +\lambda \right]~~~~,~~~~  
\lambda \equiv \left( \frac{2b}{d}\right)^{3/2}\left[ 1-\left( \frac{2b}{d}
\right)^{2} \frac{15}{32\pi b}\right] 
\label{eq:23}
\end{equation}
where$f\left( m_{0}\right) $ and $\gamma $ given by eq. (2.17). For $a=5/2,$ $N$ depends \textit{only} on the volume $V$ (linearly through $%
\gamma $) and it \textit{does not depends} on the energy $E$, and
the function $g\left( N,E,V\right) $  is 
\begin{equation}
g\left( N,E,V\right)_{a=5/2} \approx \frac{\gamma ^{N}}{(N-1)!}\left[ f\left(
m_{0}\right) +\lambda \right] ^{N-1}\frac{\lambda }{E} 
\label{eq:231}
\end{equation}

\subsection{Temperature}

From eqs.(2.9) and(2.15) the temperature for the case $a\neq 5/2$ is given by 
\[
T=\left[ \frac{\partial \left[ \ln \ \Omega _{N}(E,V)\right] }{\partial E}%
\right] ^{-1} ~~,
\]

\begin{equation}
T^{-1}-T_{0}^{-1}=8\pi Eb-\frac{3/2-a}{(5/2-a)^{2}}\left[\frac{\widehat{\gamma }\eta ( 1-\Delta ^{^{3/2-a}})^{-1}}{m_{0}^{-a}(5/2-a-\Delta ^{^{3/2-a}})^{\frac{7/2-a}{3/2-a}}}\right]^{\frac{3/2-a}{5/2-a}}E^{\frac{-1}{5/2-a}}\left[ 3/2-a+(1-\Delta ^{^{3/2-a}})^{2}\right]\ln N  
\label{eq:25}
\end{equation}
$\Delta \equiv \left( \frac{N\ m_{0}}{E}\right) $, and where we can see the
emergence of a \textit{critical temperature} 
\[
T_{0}^{-1}\equiv 4\pi d 
\]
such that the temperature satisfies $T\leq T_{0}$. We know that the
partition function $Z\left( V,T\right) $ is just the Laplace transform of
the density of states $\Omega _{N}(E,V)\,$and both are, matematically
speaking, equivalent. 
\[
Z\left( V_{0},T\right) =\int_{0}^{\infty }\Omega _{N}(m,V_{0})e^{-m/T}dm 
\]
Since $\Omega _{N}(E,V_{0})$  is of the form $e^{\frac{8\pi }{2}(d\ m+m^{2}\ d^{2})}e^{g\left( m,V_{0}\right) }$, $Z\left( V_{0},T\right) $ reads 
\[
Z\left( V_{0},T\right) =\int_{0}^{\infty }e^{-m\tau }e^{\left( \frac{m}{%
2T_{0}}\right) ^{2}}e^{g\left( m,V_{0}\right) }dm ~~~~, ~~~~\tau \equiv \frac{T_{0}-T}{TT_{0}} 
\]
and all thermodynamical functions (as the energy and pressure)
depend on the factor $(\frac{T-T_{0}}{TT_{o}})$ too. Since the exponentially increasing term for high $m$, this function diverges for all temperature reflecting the fact that the canonical frame does not exist. The function $g( m,V_{0})$ beheaves smoothly for large $m$.
\\
From eq(3.9) when $\Delta ^{^{3/2-a}}\rightarrow 0\,$, we have for the temperature:
\begin{equation}
T^{-1}-T_{0}^{-1} \approx 8\pi Eb+\frac{3/2-a}{5/2-a}\left[\frac{\widehat{
\gamma }\eta }{m_{0}^{-a}(5/2-a) ^{\frac{7/2-a}{3/2-a}}}\right]
^{\frac{3/2-a}{5/2-a}} E^{\frac{-1}{5/2-a}}\ln \left[\frac{(5/2-a) m_{0}^{-a}}{\widehat{\gamma}\eta E^{3/2-a}}\right]
\end{equation}

while when $\Delta ^{^{3/2-a}}\rightarrow \infty $, we have:
\begin{equation}
T^{-1}-T_{0}^{-1} \approx 8\pi Eb+\frac{3/2-a}{( 5/2-a) ^{2}}
E^{\frac{-1}{5/2-a}} \ln \left[\frac{m_{0}^{-a}}{\widehat{\gamma }\eta E^{3/2-a}}\right]
\end{equation}

The behaviour of the temperature as a function of the energy $E$ is as follows:

\[
\begin{tabular}{|l|l|l|}
\hline
& $T_{E\rightarrow \infty }$ & $T_{E\rightarrow 0}$ \\ \hline
&  &  \\ 
$a<3/2$ & $\approx \frac{T_{0}}{1+8\pi bET_{0}}\rightarrow \frac{1}{8\pi bE}%
=T_{H}$ & $\approx -\frac{\left( 5/2-a\right) ^{2}const\times E^{\frac{1}{%
5/2-a}}}{\left( 3/2-a\right) }\rightarrow 0_{-}$ \\ 
&  &  \\ \hline
&  &  \\ 
$a>5/2$ & $\approx \frac{T_{0}}{1+8\pi bET_{0}+const\times E^{\frac{-1}{5/2-a%
}}}\rightarrow 0_{+}$ & $\approx \frac{T_{0}}{1+const\times E^{\frac{1}{%
5/2-\left| a\right| }}T_{0}}\rightarrow T_{0}$ \\ 
&  &  \\ \hline
&  &  \\ 
$3/2<a<5/2$ & $\approx - \frac{T_{0}}{E^{\frac{1}{5/2+a}}+8\pi
bE^{\frac{7/2-a}{5/2+a}}T_{0}}\rightarrow 0_{-}$ & $\approx \frac{T_{0}}{%
1+const\times E^{\frac{-1}{5/2-a}}}\rightarrow 0_{+}$ \\ 
&  &  \\ \hline
&  &  \\ 
$a=3/2$ & $\approx \frac{T_{0}}{1+8\pi bET_{0}}\rightarrow \frac{1}{8\pi bE}%
=T_{H}$ & $\approx \frac{T_{0}}{1+8\pi bET_{0}}\rightarrow T_{0}$ \\ 
&  &  \\ \hline
\end{tabular}
\]
where\textit{\ } 
\[
T_{0}^{-1}\equiv 4\pi d 
\]
is the string temperature, and $T_{H}=\frac{1}{8\pi bE}$ which  {\it emerges} as the Hawking temperature of the system.\\

For $a=5/2$ the temperature is 
\[
T^{-1}-T_{0}^{-1}=8\pi Eb-\frac{\gamma ^{N}}{(N-1)!}\left[ f\left(
m_{0}\right) +\lambda \right] ^{N-1}\frac{\lambda }{E^{2}} 
\]
\[
T\approx \frac{T_{0}}{\left( 1+8\pi EbT_{0}\right) -\left\{ T_{0}\frac{%
\gamma ^{N}}{(N-1)!}\left[ f\left( m_{0}\right) +\lambda \right] ^{N-1}\frac{%
\lambda }{E^{2}}\right\} } 
\]

The above temperature behaviours show that the asymptotic large energy behaviour is characterized by the Hawking temperature $T_{H}$ while the small energy limit is characterized by the string temperature $ T_{0}$. Large energies corresponds to large masses and so to the classical/semiclassical behaviour of the system while small masses 
correspond to the quantum string limit.
\bigskip
Notice that for $a=5/2$, $T\rightarrow \infty $ when 
\[
T_{0}\frac{\gamma ^{N}}{(N-1)!}\left[ f\left( m_{0}\right) +\lambda \right]
^{N-1}\frac{\lambda }{E^{2}}\rightarrow \left( 1+8\pi EbT_{0}\right) \ ,%
\hspace{1cm} 
\]
that is, superheating points do appear. For low ($E\rightarrow m_{0}$) and high ($E>>m_{0}$) energies, these points are respectively:
\[
E_{low}\approx \sqrt{\left( \frac{\lambda \gamma ^{N}}{(N-1)!}\left[
f\left( m_{0}\right)+\lambda \right] ^{N-1}\right) T_{0}}~~~~~~,~~~~~~ E_{high}\approx 2\pi T_{0}\newline
\]

This is the analogue to the Carlitz situation [11] for the string gaz alone where the energy for the case $a<5/2$ is $E\simeq \left(
a+3/2\right) T_{0}^{2}/\left( T_{0}-T\right) $. This means that this case is
obviously an extremely nonuniform configuration, and the thermodynamic
properties of the various subsystems of the statistical system
under consideration are inequivalent to the properties of the system as a
whole. The microcanonical approach is sensible to the local equilibrium state, and
as a consecuence the equivalence between microcanonical and canonical
ensembles is no longer guarranteed in this case. Temperatures greater than $T_{0}$ are
allowed in the microcanonical ensemble, and the system as a whole can be
unstable, characterized by a negative specific heat. As strong interactions
are virtually present everywhere, $T_{0}$ is a universal highest temperature
for \textit{equilibrium} states. A gas of N particles initially at $T>T_{0}$
will, after some time, create other particles, and then cool down to $%
T<T_{0}$, which means that initially the gas was not in an equilibrium
state .

At energies higher than those of the superheating points, the behaviour of the temperature in the system is of the form 
\[
\left. T\right| _{E>E_{high}}\approx \frac{T_{0}}{\left( 1+8\pi
EbT_{0}\right)}\simeq \frac{1}{8\pi Eb}=T_{H} 
\]
which shows that the temperature after reaching the superheating point, begin
to have a behaviour inverse to the energy stabilyzing the system. This
behaviour shows the Hawking temperature $T_{H}$ of the system. 
This phase corresponds to the large masses and very low temperature: that is, to the semiclassical behaviour characterized by the Hawking temperature $ T_H = \frac{1}{8\pi Eb}$. In this phase, the temperature goes to zero stabilizing the system.

On the other hand, for low energies, namely $E$ lower than $E_{low}$, the behaviour of the temperature is quite different 
\[
\left. T\right| _{E<E_{low}}\approx \frac{1}{\frac{\gamma ^{N}}{(N-1)!}%
\left[ f\left( m_{0}\right) +\chi \right] ^{N-1}\frac{\chi }{E^{2}}} 
\]
which shows that before reaching the superheating point, the temperature
is proportional to the square of the energy of the system.

\subsection{Pressure and Specific Heat}

In the microcanonical ensemble, the pressure is defined by 
\[
P=T\ \ \frac{\partial \left[ \ln \ \Omega _{N}(E,V)\right] }{\partial V} 
\]
Explicitly, for $a\neq 5/2$ we have : 
\[
PV\approx \frac{T}{( 5/2-a)^{2}}\left[ \frac{\widehat{\gamma }%
\eta ( 1-\Delta ^{^{3/2-a}}) E^{3/2-a}}{( 5/2-a-\Delta
^{^{3/2-a}}) m_{0}^{-a}}\right]^{\frac{1}{5/2-a}}\ln \left[ \frac{%
( 5/2-a-\Delta ^{^{3/2-a}}) m_{0}^{-a}}{\widehat{\gamma }\eta
( 1-\Delta ^{^{3/2-a}})E^{3/2-a}}\right] 
\]
Or, as a function of $N$:
\[
\left. PV\right| _{a\neq 5/2}\approx \frac{NT}{\left( 5/2-a\right) ^{2}}\ \ln \left[
N^{-(5/2-a)}\right] = -\frac{N\ T}{\left( 5/2-a\right) }\ \ln N 
\]

The behaviour of the pressure as a function of the energy $E$ is as follows

\[
\begin{tabular}{|l|l|l|}
\hline
& $E\rightarrow 0$ & $\ E\rightarrow \infty \text{ \ \ }$ \\ \hline
&  &  \\ 
$a<3/2$ & $PV\rightarrow 0_{-}$ & $\text{\ }PV\rightarrow \infty $ \\ 
&  &  \\ \hline
&  &  \\ 
$a>5/2$ & $PV\rightarrow 0_{+}$ & $\text{\ \ }PV\rightarrow \infty $ \\ 
&  &  \\ \hline
&  &  \\ 
$3/2<a<5/2$ & $\text{\ }PV\rightarrow \infty \ $ & $\text{\ }PV\rightarrow 0$
\\ \hline
\end{tabular}
\]
and the behaviour for $a=3/2$ is: 
\[
PV\approx T\left[ \frac{\widehat{\gamma }\eta }{m_{0}^{-3/2}}\right] \ln
\left[ \frac{\ m_{0}^{-3/2}}{\widehat{\gamma }\eta \ }\right]= - N T \ \ln N  
\]
Notice that in this case $PV$ is independent of $E$.

For $a=5/2$, the expression for the pressure takes the form 
\[
\left. P\right| _{a=5/2}\approx \frac{TN}{VE}\frac{\gamma ^{N}}{(N-1)!}%
\left[ f\left( m_{0}\right) +\lambda \right] ^{N-1}\lambda 
\]
where $f\left( m_{0}\right) $ and $\gamma $ are given by eq.(2.17) and the
constant $\lambda $ is given by eq.(3.7).

We see the different behaviours of the system depending on the parameter $a$. The regimes in which the pressure becomes negative reflect the presence of gravothermal unstable behaviour, which is accompassed, as we see below, by the behaviour of the specific 
heat.

\bigskip

Let us discuss the specific heat $ C_{v}=\left. \frac{\partial E}{\partial T}\right| _{V=fixed}$.   From eq.(3.9) it takes for our system the explicit form

\[
C_{v}\approx \frac{-1}{T^{2}}\left\{ \frac{1}{8\pi b-X\left[ \frac{X}{N}+\ln
N\ E^{7/2-a}Y\left( \Delta \right) \right] }\right\} 
\]
where  
\[
X\equiv \frac{\left[ \left( T_{0}^{-1}+8\pi bE\right) -T^{-1}\right] }{\ln N}%
\left( 5/2-a\right) ~~~~;\hspace{2.57cm}\Delta \equiv \left( \frac{N\ m_{0}}{E}%
\right) 
\]
\begin{equation}
Y\left(\Delta \right) \equiv \frac{\left( 3/2-a\right) \Delta
^{^{3/2-a}}\left( 1-\Delta ^{^{3/2-a}}\right)}{\left[ 3/2-a+\left( 1-\Delta
^{^{3/2-a}}\right) ^{2}\right]} - \left[ 1+\frac{\left( 3/2-a\right) \Delta ^{^{3/2-a}}}{5/2-a}
\left( \frac{3/2-a}{\left( 1-\Delta ^{^{3/2-a}}\right) }+\frac{7/2-a}{\left(
5/2 - a -\Delta ^{^{3/2+a}}\right) ^{\frac{7/2-a}{3/2-a}}}\right)\right]
\end{equation}

The behaviour of the specific heat as a function of the energy $E$ is as
follows:

\[
\begin{tabular}{|l|l|l|}
\hline
& $E\rightarrow 0\text{ }$ & $E\rightarrow \infty \text{ \ }$ \\ \hline
&  &  \\ 
$a<3/2$ & $C_{v}\rightarrow \ 0_{-}\ $ & $\text{\ }C_{v}\rightarrow \frac{-1}{%
8\pi bT^{2}}\approx const\ $ \\ 
&  &  \\ \hline
&  &  \\ 
$a>5/2$ & $\text{\ \ \ }C_{v}\rightarrow \frac{-1}{8\pi bT^{2}}\approx const$
& $C_{v}\rightarrow \ 0_{-}$ \\ 
&  &  \\ \hline
&  &  \\ 
$3/2<a<5/2$ & $\text{\ }C_{v}\rightarrow \ 0_{-}\ \ $ & $\text{\ \ }%
C_{v}\rightarrow \frac{-1}{8\pi bT^{2}}\approx const$ \\ \hline
\end{tabular}
\]

For $a=5/2$, $C_{v}$ takes the form

\[
C_{v}\approx -\frac{E^{3}}{T^{2}\left[ 8\pi E^{3}\ b+2\frac{\gamma ^{N}}{%
(N-1)!}\left[ f\left( m_{0}\right) +\lambda \right] ^{N-1}\lambda \right] } 
\]
with the following behaviours
\[
\left. C_{v}\right. _{E\rightarrow \infty }\rightarrow \frac{-1}{8\pi bT^{2}}%
\ \approx const\ \ \ ;\ \ \ \ \ \left. C_{v}\right. _{E\rightarrow
0}\rightarrow \ 0_{-} 
\]
Notice that the changes in the local energy density given by the
microcanonical ensemble generate an instability that is here directely
translated into the negative value for the specific heat $C_{v}$.
The canonical ensemble (which takes the average of the energy) does not take
care of the local fluctuations of the energy density and, for instance, both
approaches (canonical and microcanonical) are not equivalent in this case.

\section{Duality (crossing) of the scattering amplitudes}

A characteristic property common to string systems and dual models is the duality
of the scattering amplitudes (crossing symmetry). This means that the four
point amplitude can be expressed as a sum over resonances either in the $s $ or
$t$ channel, even at very high energies, ($s$,$t$ being the center of mass energy and the transverse momentum of the components). As was pointed out in ref.[10], in
order for duality be valid, the number of $n$-body channels open in the
statistical model, and so the total number of open channels, must rise in
parallel with the number of resonances as the center of mass energy is
increased 
\[
N_{n}\left( m\right) \sim \rho _{string}\left( m\right) \ \sim d\ m^{-a}e^{%
\frac{8\pi }{2}d\ m}\ ,\ \ \ \ \ \ m\rightarrow \infty 
\]
where $N_{n}\left( m\right) $ is the number of open $n$-body channels at the
center of mass energy $m$. In our system of black-holes and strings, we expect 
$N_{n}\left( m\right) \sim \rho \left( m\right) $, with $\rho \left( m\right) $ as given by eq.(2.2).
An explicit expression for the two-body amplitude is
\begin{equation}
N_{n}\left( m\right) =\frac{1}{2!}\int_{m_{0}}^{m-m_{0}}\rho \left(
m_{2}\right) dm_{2}\int_{m_{0}}^{m-m_{2}}\rho \left( m_{1}\right) dm_{1} 
\label{eq:26}
\end{equation}
If duality of the scattering amplitude can be argued to be a symmetry of the
string-black-hole system, one should support it with a direct computation of $
N_{n}\left( m\right) $ eq. (4.1) by using $\rho \left( m\right) $ given by
eq. (2.2). If we assume the lowest mass of the system $m_{0}=0$, we obtain 
\[
N_{2}\left( m\right) =\frac{c^{2}}{2!}\int_{0}^{m}dm_{2}%
\int_{0}^{m-m_{2}}dm_{1}e^{r_{1}+r_{2}} 
\]
where 
\[
r_{i}\equiv -a\ \log \left[ \ m_{i}(d + m_{i}b)\right] + 4\pi \text{ }%
m_{i} (d + m_{i}b) 
\]
and $b, d $ as fixed by eq.(2.22). We can easily see,
that the dominant contribution is obtained when $m\gg m_{2}$ and $m\simeq
m_{1}$. This is similar to the evaluation of the density of states with
$n$-strings, in which most of the energy is carried by one body of the system
and the $n-1$ others share the ligth remnants. We obtain in this approximation, 
\begin{equation}
N_{2}(m)\approx \frac{c^{2}}{2}\left( d\text{ }m\ +\text{ }b\text{ }%
m^{2}\right) ^{-a}\text{ }e^{\frac{8\pi }{2}(d\ m+ b m^{2}\ )}\ \ \ ,\ \ \
\ \ m\rightarrow \infty  \label{eq:27}
\end{equation}
which is precisely the degeneracy of states $\rho \left( m \right)
$ of the string-black hole system in which the exponential part dominates. 
That means that this argument can be extended to any $n$, and
we find for our string-black hole system, the same result as in the string
and dual models: the number of open channels does grow in parallel with the
degeneracy of states, here given by eq.(4.2), as energy is increased.

\section{Microcanonical Field formulation: scattering amplitudes and the
string/particle-black hole system}

Is clear from the previous sections that the correct thermodynamical
interpretation of the gaz of black-holes and strings is in the
microcanonical formulation. Let us go further in the understanding of this system and study a field theory description within the microcanonical field (or thermo field) formulation for this system. The physical component of the propagator in the microcanonical field formulation [15] is given by 
\begin{equation}
\Delta _{E}^{11}\left( k\right) =\frac{1}{k^{2}-m^{2}+i\epsilon }-2\pi
i\delta \left( k^{2}-m^{2}\right) n_{E}\left( m,k\right) \label{eq:28}
\end{equation}
where the second term corresponds to the statistical
(microcanonical) part and $n_{E}\left( m,k\right)$ is the microcanonical number density:
\[
n_{E}\left( m,k\right) =\sum _{l=1}^{\infty }\frac{%
\Omega \left( E-l\omega _{k}\left( m\right) \right) }{\Omega \left( E\right) 
}\theta \left( E-l\omega _{k}\right), 
\]
 $l,\omega _{k}$ and $\theta
\left( E-l\omega _{k}\right) $ being the mode number, the dispersion relation
and the step function respectively. In our case, $n_{E}\left( m,k\right)$ is explicitly given
by 
\[ 
n_{E}\left( m,k\right) =\sum _{l=1}^{\infty }\frac{%
c\left[ d^{2}\left( E-l\omega _{k}\left( m\right) \right) ^{2}+d\left(
E-l\omega _{k}\left( m\right) \right) \right] ^{-a}e^{4\pi d\left( E-l\omega
_{k}\right) }e^{4\pi d^{2}\left( E-l\omega _{k}\right) ^{2}}}{c\left(
dE+d^{2}E^{2}\right) ^{-a}e^{4\pi dE}e^{4\pi d^{2}E^{2}}} \theta \left( E-l\omega _{k}\right)
\]
$\allowbreak $or (with $E=M$, being $M$ the total mass of the system):
\[
n_{M}=\sum _{l=1}^{M/\omega _{k}} \left( 1-\frac{%
l\omega _{k}d}{(Md+1)}\right) ^{-a}\left( 1-\frac{l\omega _{k}}{M}\right)
^{-a}e^{-4\pi dl\omega _{k}}e^{-8\pi d^{2}Ml\omega _{k}}e^{4\pi \left(
dl\omega _{k}\right) ^{2}} 
\]
Let us customize the above expression for the number
density as 
\begin{equation}
n_{M}=\sum _{l=1}^{M/\omega _{k}} \left[ -\frac{l\omega
_{k}}{M} + \sum _{n=0}^{\infty } \frac{\left( M\varsigma
\right) ^{n}}{n!}\left( 1-\frac{l\omega _{k}\varsigma }{n+1}\right) \right]
^{-a}\left( 1-\frac{l\omega _{k}d}{(Md+1)}\right) ^{-a}e^{4\pi \left[ \left(
dl\omega _{k}\right) ^{2}-l\omega _{k}d\right] } \label{eq:29}
\end{equation}
where 
\[
\varsigma \equiv \frac{8\pi d^{2}l\omega _{k}}{a} 
\]
In order to obtain a complete picture of the string-black-hole system, it is
instructive to compute explicitly from the microcanonical statistical
propagator eq.(5.1) the four point transition amplitude  $A$. To do this, it is
necessary to insert in eq.(5.1) the expression eq.(5.2) for $n_{M}$ and use the
relation between the propagator and the four point transition amplitude process given
explicitly in the time ordered perturbation theory. Thus, we find for the total amplitude:
\begin{eqnarray}
A_{total} &=&A_{0} +A_{str-bh}~,  \label{eq:291} \\
A_{0} &\equiv &\left. \frac{\lambda _{0}^{2}}{4\pi ^{2}i}\frac{m^{2}\delta
\left( p_{1}^{\prime }+p_{2}^{\prime }-p_{1}-p_{2}\right) }{\sqrt{%
\varepsilon _{1}^{\prime }\varepsilon _{2}^{\prime }\varepsilon
_{1}\varepsilon _{2}}}\left[ \frac{1}{\left( p_{1}^{\prime }-p_{1}\right)
^{2}-m^{2}+i\epsilon }+\frac{1}{\left( p_{2}^{\prime }-p_{1}\right)
^{2}-m^{2}+i\epsilon }\right], \right.  \nonumber \\
A_{str-bh} &\equiv &-2\pi i\sum_{l=1}^{M/\omega _{\left( m\right)}}\left[ -\frac{l\omega _{\left( m\right) }}{M}%
+ \sum_{n=0}^{\infty }\frac{\left( M\varsigma \right)
^{n}}{n!}\left( 1-\frac{l\omega _{\left( m\right) }\varsigma }{n+1}\right)
\right] ^{-a}\left( 1-\frac{l\omega_{\left( m\right)}d}{(Md+1)}\right) ^{-a}e^{\varkappa } \label{eq:30}
\end{eqnarray}
with \[
\varkappa \equiv 4\pi \left[ \left( dl\omega _{\left( m\right) }\right)
^{2}-l\omega _{\left( m\right) }d\right] 
\]
$\lambda _{0}$ being the coupling constant of the scattering amplitude and
the energy condition is on mass-shell 
\[
\varepsilon _{1}^{\prime }\equiv p_{0}=\sqrt{m^{2}+\overline{p}^{2}} 
\]
As is known, the transition amplitude $A$ (in the momentum representation) between an
initial state $\left| i\right\rangle $ and  final state $\left|
f\right\rangle $ is defined from the $S$ matrix  $\left(
S=1+iT\right)$ as  $\left\langle f\left| T\right| i\right\rangle =\left( 2\pi \right) ^{4}\delta \left( P^{\prime }-P\right) \left\langle f\left| A\right| i\right\rangle$,
 $P^{\prime },P$ being the initial and final four-momenta. In order to
establish and show the kind of relation between the string-particle part and the
black-hole part of the spectrum, let us use the  formal
correspondence [18] between the Feynman propagator $\Delta _{ij}$ and the
expression for dual string amplitudes as [19]:
\[
\Delta _{ij}=[s_{ij}+\alpha ^{\prime }M^{2}+\alpha \left( 0\right) ]^{-1} 
\]
where $s_{ij}=\left( p_{i}+p_{i+1}+....+p_{j}\right) ^{2}$ ; $\alpha \left(
0\right) $ is the Regge trayectory at zero momentum; $\alpha \left(
s_{ij}\right) =\alpha \left( 0\right) +\alpha ^{\prime }s_{ij}$ and $\alpha
^{\prime }M^{2}=\sum_{n=1}^{\infty }na_{n\mu }^{+}a_{n}^{\mu }$ \ is the
usual mass operator. With these definitions, we can see that the expression eq.(5.4)
contains the full spectrum of the system: the known string-particle spectrum
and on the other hand, the {\bf new} part $A_{str-bh}$ describing the string-black hole component of the spectrum : 
\begin{equation}
A \approx A_{str-particle}+A_{str-bh}~,  
\end{equation}
\label{eq:292} 
\begin{equation}
A_{str-particle} \equiv \frac{\lambda ^{2}}{2\pi^{2}}
{\sum_{n=0}^{\infty }}\frac{\Gamma \left(n+1+\alpha \left(t\right)
\right)}{n!\Gamma \left(1+\alpha \left( t\right) \right)}\frac{1}{
n+\alpha \left( s\right)}     
\end{equation}
\label{eq:293} 

here $s,t$ are the channels of the process in the Mandelstam formulation
and we used the well known relation between the Euler Gamma function and the
integral representation of the Veneziano amplitude 
\[
A_{str-particle}=\int dx\text{ }x^{p_{1}.p_{^{1^{\prime }}}}(1-x)^{p_{1^{^{\prime
}}}.p_{2}}=\frac{\lambda ^{2}}{2\pi ^{2}}
\sum_{n=0}^{\infty } \frac{\Gamma \left( n+1+\alpha \left( t\right) \right) }{n!\Gamma
\left( 1 + \alpha \left( t\right) \right) }\frac{1}{n + \alpha \left( s\right) } 
\]
On the other hand, the microcanonical statistical part $A_{str-bh}$ given explicitely by eq.(5.4) describes the string-black hole component, and we can easily see that in the limit 
$l\rightarrow M/\omega \left( m\right) $ the exponentially increasing mass
density of states for the string-black-hole  system is recovered: 
\[
A_{str-bh}\approx \left[( Md )^2 + Md \right]^{-a}e^{4\pi
\left[ \left( M d \right)^{2} + M d\right] }\cong \rho \left(
m\right) 
\]

We analize the mass spectrum of the system in section VI below.

\subsection{Microcanonical Partition function from the propagators and scattering amplitudes }

Now, we can derive the partition function in the 
microcanonical field approach from the propagators and the scattering amplitudes computed
above. The relation between the S matrix formulation in Quantum Field Theory and the
statistical operator in the canonical ensemble [16,20] can be easily extended
to the microcanonical description as follows. The total propagator operator (with
its statistical microcanonical part $G(E)$) 
\begin{equation}
\widehat{G}\left( E\right) =\frac{1}{E-\widehat{H}-i\epsilon }+ G(E),
\label{eq:31}
\end{equation}
 and the S-matrix are related as 
\[
\mathbb{I}_{m}\widehat{G}\left( E\right) =\mathbb{I}_{m}\widehat{G}_{0}\left(
E\right) +\frac{1}{4i}\widehat{S}^{-1}\left( E\right) \frac{%
\overleftrightarrow{\partial }}{\partial E}\widehat{S}\left( E\right) 
\]
where $\mathbb{I}_{m}$ stands for the imaginary part,  $\widehat{S}\left( E\right) $ is the scattering operator at the energy 
$E$ and $\widehat{G}_{0}\left( E\right) $ is the free part of the full
propagator (which yields non-conected graphs in the cluster expansion). As is known, the relation to the physical $\widehat{T}\left( E\right)$ scattering matrix is
$\widehat{S}\left( E\right) =1+i\delta \left( E-\widehat{H}_{0}\right) 
\widehat{T}\left( E\right)$, 
$\widehat{H}_{0}$ being the free Hamiltonian. The logarithm of the trace of the canonical partition function $Z\left( T,V\right)$ can be written as [16] 
\[
\ln Z\left(T,V\right) =\ln Z_{0}\left(T,V\right) + Tr \int d^{4}P e^{-\overline{b}.
P}\frac{-1}{\pi }\delta ^{3}\left(\overline{P}-\widehat{
\overline{P}}\right) \mathbb{I}_{m}\left[\widehat{G}\left( E\right) -\widehat{G}_{0}\left(E\right) \right] _{connected} 
\]
where the four-vector temperature $\overline{b}^{\mu }$ is defined by the
identity $ \overline{b}^{\mu }=\frac{1}{T}u^{\mu } $, 
$T$ being the temperature in the rest frame of the box enclosing the
thermodynamical system, $u^{\mu }u_{\mu }=1\ $being its four-velocity. The
index $connected$ means that only the connected part of the full propagator $%
\widehat{G}\left( E\right) $ must be taken, which is simply done by substracting from the full propagator the free part $\widehat{G}_{0}\left( E\right) $ corresponding to
an ideal gas configuration. Recalling that the operation of taking the connected part leaves the operators invariant, we can write 
\[
\ln Z\left( T,V\right) =\ln Z_{0}\left( T,V\right) +\int e^{-\overline{b}%
.P}d^{4}P\ \rho _{I}\left( P^{2},\mathbb{V}.P,\mathbb{V}^{2}\right)~~,~~ \mathbb{V}^{\mu }\mathbb{\equiv }\frac{2V}{\left( 2\pi \right) ^{3}}u^{\mu }\ \ 
\text{(four-vector volume)}
\]
where, by analogy with the interaction level density, the function which we
call mass spectrum, or cluster level density, is given by 
\begin{equation}
\rho _{I}\left( P^{2},\mathbb{V}.P,\mathbb{V}^{2}\right) =\left\{ \frac{-1}{\pi }\delta \left( \overline{P}-\widehat{\overline{P_{0}}}\right) \mathbb{I}_{m}\left[ \widehat{G}%
\left( E\right) -\widehat{G}_{0}\left( E\right) \right] \right\} _{connected}
\label{eq:32}
\end{equation}
On the other hand, from the pure statistical point of view: 
\begin{equation}
\rho _{I}\left(P^{2},\mathbb{V}.P,\mathbb{V}^{2}\right) =\rho -\rho _{0}  
\label{eq:33}
\end{equation}
$\rho $ being the full density of states of the system under consideration
and $\rho _{0}$ the density of states in the free configuration,
\begin{equation}
\rho _{0}=\sum_{k=1}^{\infty } \mathbb{V}^{\mu }P_{\mu}\delta \left( P^{2}-k^{2}m^{2}\right).  \label{eq:34}
\end{equation}
$\rho $ satisfies the following relation 
\begin{equation}
\rho \left( P^{2},\mathbb{V}.P,\mathbb{V}^{2}\right) =\mathbb{V}^{\mu }P_{\mu }\text{ }\rho
\left( \sqrt{P^{2}}\right)  
\label{eq:35}
\end{equation}
This condition means that the cluster counting $\rho
\left( P^{2},\mathbb{V}.P,\mathbb{V}^{2}\right)$ in the rest frame of the system can be
reexpressed as the counting of states for a single particle of the mass degeneracy $\rho
\left( m\right) $ moving in the volume $V$. Now, since the
mathematical mapping between the microcanonical and canonical ensembles, is
easy to see that the partition function in the microcanonical ensemble we are looking for is given by
\begin{equation}
\ln \Omega \left( E,V\right) =\ln \Omega _{0}\left( E,V \right) + Tr\int
d^{4}P\left\{ \frac{-1}{\pi }\delta ^{3}\left( \overline{P}-\widehat{%
\overline{P}}\right) \mathbb{I}_{m}\left[ \widehat{G}\left( E\right) -\widehat{G%
}_{0}\left( E\right) \right] \right\} _{connected} 
\end{equation}
Considering that the imaginary part $\mathbb{I}_{m}\left[ \widehat{G}\left(
E\right) -\widehat{G}_{0}\left( E\right) \right] $ is the connected part of
the full propagator $\widehat{G}\left( E\right)$, precisely corresponds
to the physical component $\Delta
_{E}^{11}\left( k\right)$ eq.(5.1)of the microcanonical field formulation which leads to the scattering amplitude expresion eq.(5.5).
From the comparison between the propagator eq.(5.1), (5.2) with the
expressions eq.(5.7),(5.8) yielding the level density of states $\rho _{I\text{ }%
}$, we can easily see that the dynamical
information from the connected part of the propagator eq.(5.8) is automatically
translated into a variation of the energy (mass) levels in the statistical
ensemble (given by $\rho \left( m\right))$ and the statistical information
of the propagator eq.(5.1).

\bigskip
We will consider now the system of black holes and strings as N-body extended objects in a Nambu-Goto formulation and derive the microcanonical propagator and partition function  from the Nambu-Goto action of N-body extended objects and from them we will obtain the mass spectrum of the system.

\section{The Nambu-Goto action and the string-particle-black hole spectrum}

It is difficult to study this system in the Hamiltonian framework because of
the constraints and the vanishing of the Hamiltonian. As is known, the
Nambu-Goto action\ is invariant under the reparametrizations 
\[
\tau \rightarrow \widetilde{\tau }=f_{1}\left( \tau ,\sigma \right) \text{ }%
\ \ \sigma \rightarrow \widetilde{\sigma }=f_{2}\left( \tau ,\sigma \right) 
\]
then, we can make the following choice for the dynamical variable $x_{0}$ and
the space variable $x_{1}$, as first proposed in ref.[21] which does not restrict the essential physics and simplifies considerably the dynamics of the system 
\[
x_{0}\left( \tau ,\sigma \right) \equiv x_{0}\left( \tau \right) ;\ \ \
x_{1}\left( \tau ,\sigma \right) \equiv \kappa \sigma \text{ \ \ \ \ }\left(
\kappa =const\right) 
\]
For this, it is sufficient to make the chain derivatives and to write
the action in the form 
\[
S=-\frac{\kappa }{\alpha ^{\prime }}\int_{\tau 1}^{\tau 2}\stackrel{.}{x}%
_{0}d\sigma \ d\tau \ \sqrt{\left[ 1-\left( \partial _{0}x_{b}\right)
^{2}\right] \left[ 1+\left( \partial _{1}x_{a}\right) ^{2}\right] }~,
\]
where $a,b=2,3;\ \partial _{1}x_{a}=\varepsilon _{1a}^{\ \ \
0b}\partial _{0}x_{b}$ and in order to simplify at maximum this action
we choose an orthonormal frame; (thus passing from the Nambu-Goto action to
the Born-Infeld representation). Therefore, the invariance with respect to the choice of the coordinate evolution parameter means that one of the dynamical variables of the theory ($x_{0}\left( \tau \right)$ in this case) becomes the observed time with the corresponding non-zero Hamiltonian 
\begin{equation}
H_{BI}=\Pi _{a}\stackrel{.}{x}^{a}-L=\sqrt{\alpha ^{2}-\Pi _{b}\Pi ^{b}}~~,~~ 
 \Pi ^{b}=\frac{\partial L}{\partial \left( \partial _{0}x_{b}\right) }%
\hspace{0.5cm},\hspace{1cm}\alpha \equiv \frac{\kappa \sqrt{1+\left(
\partial _{1}x_{a}\right) ^{2}}}{\alpha ^{\prime }} 
\label{eq:36}
\end{equation}

Now, in order to find the microcanonical partition function of the system from the Nambu Goto Hamiltonian we proceed
as follows: From the most simple quantum path-integral formalism, we have 
\[
K\left( q^{\prime },t,q,0\right) \equiv \left\langle q^{\prime }\left|
\left( e^{H\varepsilon }\right) ^{N}\right| q\right\rangle =\left\langle
q^{\prime }\left| \Psi \left( r,s,t...\right) \right. \right\rangle 
\]
where $K\left( q^{\prime },t,q,0\right) $ is the propagator, $H$ is the
Hamiltonian, $t$ is the time that was fractionated in small
lapses $t=N\varepsilon $ and $q,q^{\prime }$ and $\Psi \left( r,s,...\right) 
$ are the physical states with $r,s...$ quantum numbers. With the usual path
integral operations and introducing the integral representation for a
pseudodifferential operator [17] 
\[
\int \left( t^{2}+u^{2}\right) ^{-\lambda }e^{itx}dt=\frac{2\pi ^{1/2}}{%
\Gamma \left( \lambda \right) }\left( \frac{\left| x\right| }{2u}\right)
^{\lambda -1/2}K_{\lambda -1/2}\left( u\left| x\right| \right) 
\]
where $K_{\nu }\left( x\right) $ is the Mac Donald's function, the
propagator for a sub-interval takes the form 
\[
K_{q_{j},q_{j+1}}=\delta _{q_{j},q_{j+1}}-i\varepsilon \left[ 4\alpha \frac{%
K_{-1}\left( \alpha \left| q_{j}-q_{j+1}\right| \right) }{\left|
q_{j}-q_{j+1}\right| }\right] 
\]

Putting all the subinterval propagators together yields the full 
propagator 
\[
K=\delta _{q_{N},q_{0}}-iN\varepsilon \left[ 4\alpha \frac{K_{-1}\left(
\alpha \left| q_{N}-q_{0}\right| \right) }{\left| q_{N}-q_{0}\right| }%
\right] 
\]

Making, without lost of generality, the transformation $-it \rightarrow -\beta $
, integrating and Fourier transforming to momentum space,
yields the canonical partition function
\begin{equation}
Z = \sum_{N}\left[1 - \beta 4\alpha \frac{K_{-1}(\alpha |q_{N}- q_{0}|)} {|q_{N}- q_{0}|}\right] = \sum_{N}\exp -\beta \left[ 4\alpha \frac{
K_{-1}(\alpha |q_{N}-q_{0}|)}{|q_{N}-q_{0}|}\right]
\label{eq:37}
\end{equation}
The microcanonical partition function $\Omega _{m}$ is obtained as the
inverse Laplace transform of the last expression : 
\[
\Omega _{m}=\delta \left( E\right) -i\sum_{N=1}^{\infty } \sum_{\overline{p}_{1} }\sum_{n_{1}=1}^{\infty } 
.......\sum_{\overline{p}_{N}} \sum_{ n_{n}=1}^{\infty }  \left[ 4\alpha \frac{K_{-1}\left( \alpha \left|
\sum n_{j}\varepsilon _{j}-E\right| \right) }{E^{2}}\frac{1}{
n_{1}n_{2}.....n_{N}}\right] 
\]
where the factor $\frac{1}{n_{1}n_{2}.....n_{N}}$ allows eliminate the
overcounting, and $\sum n_{j}\varepsilon _{j}=E_{N}$.
\bigskip

The microcanonical propagator for the string-particle-black hole system
can be consistently formulated using the relation between time ordered products and normal products 
\[
-iT\left[ \varphi \left( x\right) \varphi \left( 0\right) \right]
=D_{F}\left( x\right) -i:\varphi \left( x\right) \varphi \left( 0\right) : 
\]
where $D_{F}\left( x\right)
=-i\left\langle \mathcal{J}\left| \varphi \left( x\right) \varphi \left(
0\right) \right| \mathcal{J}\right\rangle $ is the ordinary Feynman
propagator with the expectation value evaluated in the basic states of our
system ( i.e. for zero temperature ) 
\[
\left| \mathcal{J}\right\rangle =\left[ {k,m}{\prod }{
n_{k},m}{\sum }\right] {k,m}{\prod }\left| n_{k,m}\right\rangle
\otimes \left| \widetilde{n}_{k,m}\right\rangle 
\]
Since the relation between microcanonical and canonical formulations is through
a Laplace transform, it is reasonable to perform the following mapping 
\begin{equation}
\int_{0}^{\infty }dE^{\prime }\Omega _{E-E^{\prime }}D_{E^{\prime }}= D_{F}\Omega _{E}-i\sum_{N=1}^{\infty } \sum_{\overline{p}_{1}}\sum_{n_{1}=1}^{\infty }
.......\sum_{\overline{p}_{N}} \sum_{ n_{n}=1}^{\infty }\left[ 4\alpha \frac{K_{-1}\left( \alpha \left| \sum
n_{j}\varepsilon _{j}-E\right| \right) }{E^{2}}\frac{1}{n_{1}n_{2}.....n_{N}}%
\right] \left\langle \mathcal{J}\right| :\varphi \left( x\right) \varphi
\left( 0\right):\left| E\right\rangle  
\label{eq:38}
\end{equation}

where $D_{E^{\prime }}$ is the
microcanonical propagator and we defined the microcanonical string/particle-black hole state as 
\[
\left| E\right\rangle =\frac{1}{\Omega _{E}}\int_{0}^{\infty }dE^{\prime
}\Omega _{E-E^{\prime }}L_{E^{\prime }}^{-1}\left[ \left| \beta
\right\rangle \right] 
\]
L$^{-1}$ being the inverse Laplace transform. The matrix element for the most general states in our system is 
\[
\left\langle \mathcal{J}\right| :\varphi \left( x\right) \varphi \left(
0\right) :\left| E\right\rangle =\frac{1}{\Omega _{E}}\Omega _{E-E^{\prime }}
{\sum_{\overline{p}} }\left[ \frac{n}{\varepsilon _{p}V}\cos
\left( \overline{p}.\overline{x}-\varepsilon _{p}t\right) \right] 
\]
By inserting this expression into the definition of the microcanonical
propagator $D_{E^{\prime }}$ eq.(6.3) given above and converting the momentum
sum into an integral, we have 
\[
D_{E}(t,\overline{x})=\delta \left( E\right) D_{F}-4i\alpha \int \frac{d^{3}p}{\left( 2\pi \right) ^{3}\varepsilon _{j}}\sum_{n=1}^{\infty} \frac{K_{-1}\left( \alpha \left| n_{j}\varepsilon_{j}-E\right| \right)}{E^{2}}\frac{\Omega \left( E-n_{j}\varepsilon
_{j}\right) }{\Omega \left( E\right)}\cos \left( \overline{p}.\overline{x}
-\varepsilon _{j}t\right) 
\]
Finally, Fourier transform to momentum representation gives the microcanonical propagator of the system:
\begin{equation}
D_{E}(k,m) =\frac{\delta \left(E\right)}{\omega^{2}-k^{2}-m^{2}+i\varepsilon }- D_{E~(str-bh)}(k,m)
\label{eq:381}
\end{equation}

\begin{equation}
D_{E~(str-bh)}(k,m)= 8\pi i\alpha \delta \left( \omega ^{2}-k^{2}-m^{2}\right) \sum_{l=1}^{\infty } \frac{K_{-1}\left( \alpha \left|
l\omega _{k}-E\right| \right) }{E^{2}}\frac{\Omega \left( E-l\omega
_{k}\right)}{\Omega \left( E\right)}\theta \left(E-l\omega _{k}\right) 
\end{equation}

where $\theta \left( x\right) $ is the usual step function and $K_{-1}$ is the Mac Donald's function. The first term in the microcanonical propagator is the usual Feynman
propagator, the second one is the new microcanonical part $D_{E~(str-bh)}(k,m)$. This
part is crucial for the correct description of the full string-particle-black hole system, as we can see by explicitly expanding the Mac Donald's function $K_{-1}$ in the second term of the microcanonical propagator eq.(6.4) and expressing it as a function of the mass, being $E=M$ the total mass of the system:
\begin{equation}
D_{E~(str-bh)}(k,m) = 8\pi i\alpha \delta \left( \omega ^{2}-k^{2}-m^{2}\right) \sum_{l=1}^{M/\omega _{k}}  \ln \left( \frac{\gamma _{e}\alpha
\left| l\omega _{k}-M\right| }{2}\right) \frac{\alpha \left| l\omega
_{k}-M\right| }{2} \sum_{s=0}^{\infty} \frac{\left(
\alpha \left| l\omega _{k}-M\right| /2\right) ^{2s}}{s\Gamma \left(
s+2\right)}-  \nonumber
\end{equation}

\begin{equation}
-\frac{1}{2} \sum_{s=0}^{\infty} \frac{\left(\alpha \left| l\omega _{k}-M\right| /2\right) ^{2s+1}}{s!\Gamma \left(s\right)}\left(\sum_{h=1}^{s}\frac{1}{h}+
\sum_{h=1}^{s+1} \frac{1}{h}\right) +\frac{\alpha
\left| l\omega _{k}-M\right| }{4} \frac{\theta \left( M-l\omega
{k}\right)}{M^{2}}\frac{\Omega \left( M-l\omega _{k}\right)}{\Omega
\left(M\right)}
\label{eq:391}
\end{equation}

We see that when $M\rightarrow 0$ this expression yields the pure
string-like behaviour (Gamma type string-amplitude). When $M$ grows it pass
for all the intermediate states up to the pure black-hole behaviour. Notice that,
previously [18],[19], the relation between the Feynman propagator and the
Veneziano amplitude was putted "by hand". We see that this type of
structure coming from the statistical microcanonical part is contained in
our microcanonical propagator eq.(6.5). Notice that the relation between
temporal and normal ordering of the field operators contributes to the
statistical part of this full propagator. 
\bigskip
Explicitly, for $a>0$ the spectrum
(singular points of the microcanonical part) is given by the following
expression 
\begin{equation}
\sum_{l=1}^{M/\omega _{k}}\sum_{n=0}^{\infty} \frac{1}{n!}\left[\left( \frac{8\pi l\omega
_{k}d^{2}M}{a}\right)^{n}\left( 1-\frac{8\pi \left( l\omega _{k}d\right)
^{2}}{a\left( n+1\right) }\right) \right] =\sum_{l=1}^{M/\omega _{k}} \frac{l\omega _{k}}{M} 
\label{eq:392}
\end{equation}

It is interesting to note that the main difference between our
microcanonical propagator eq.(6.4) and the propagator $\Delta _{11}$ of the previous
section is in that the propagator eq.(6.4) includes all non-local effects:
from the $N$-bodies of the system as extended objects (i.e. strings) and from
the derivation of this propagator from a theory with a Hamiltonian not
quadratic in the momenta as in the Nambu-Goto formulation of string theory.
For instance, the propagator eq.(6.4) becomes the propagator eq.(5.1) when the
extended bodies became point-particles and the Hamiltonian is quadratic in the momenta
(constrained particle Hamiltonian. Let us analyze the excited regime $l\rightarrow
M/\omega _{k}$. 

\bigskip

The microcanonical part of the propagator eq.(6.4) in the limit $l\rightarrow
M/\omega _{k}$ takes the form 
\begin{equation}
D_{E~(str-bh)}(k,m)= 8\pi i\alpha \delta \left(\omega ^{2}- k^{2}- m^{2}\right) \frac{1}{2M^{2}}
 \left\{2e^{(\frac{8\pi dM}{a})^{2}}
\left[\sum_{n=0}^{\infty}\left(\frac{a}{8\pi}\right)^{n} \left(1 -\frac{8\pi (Md)^{2}}{a(
n+1)}\right)\right]-1\right\}^{-a} \times 
\nonumber
\end{equation}
\begin{equation}
\left(\frac{1}{Md+1}\right)^{-a}
 \times e^{4\pi Md(Md-1)}\left[\lim_{l\rightarrow M/\omega _{k}} \frac{1}{|l\omega _{k}-M|}\right],
\end {equation}
and the spectrum is given by 
\[
e^{-\left( \frac{8\pi dM}{a}\right)^{2}}=2\sum_{n=0}^{\infty }\left[ \left( \frac{a}{8\pi }\right) ^{n}\left( 1-\frac{8\pi \left(
Md\right)^{2}}{a\left( n+1\right) }\right) \right] 
\]
For $n>>1$ the spectrum becomes 
\begin{equation}
\left( \frac{8\pi d}{a}M\right) ^{2}\simeq n\ln (\frac{8\pi}{a}) -\ln 2
\label{eq:40}
\end{equation}

This is the spectrum of the system in the very quantum regime, and we see that this is like the pure quantum string spectrum. That is, quantum black holes have their mass quantized as quantum strings.

\section{Conclusions}

We have first considered the ideal gaz of black holes and strings in a microcanonical formulation and analyzed the microcanonical content of this system (sections II and III). We then considered the system with interactions, first in a quantum field theory microcanonical formulation (sections IV,V), second as N-body extended objects in a Nambu Goto formulation, from which we found the propagator, scattering amplitudes and mass spectrum.
This study allowed us to describe a wide class of properties and physical magnitudes of the system, covering the different mass ranges and the classical, semiclassical and quantum behaviours of the system.\\

In the gaz of strings and black-holes, the global behaviour of the string-particle-black hole  system turns to be divided in two main cases: $a=5/2$ and $a\neq 5/2$, but several relevant features are generic, common to both cases, as the fact that the Hawking temperature and the string temperature emerge in the asymptotic large energy regime and in the small energy regime respectively. We found that the temperature of the system is given by 
\[
T= \frac{T_0}{1 + 8 \pi E bT_0 + F(E)T_0}
\]
with $F(E)/E \rightarrow 0$ for $E\rightarrow \infty$ and $F(E) \rightarrow 0$ for $E\rightarrow 0$. The function $F(E)$ has been explicitely obtained in the two different cases of interest: $a=5/2$ and $a\neq 5/2$. \newline

For $a=5/2$, the number of components $N$ \textit{does not} depends on the energy $E$, but depends on the minimum mass of the system $m_{0}$. 
The microcanonical density of states $\Omega _{N}(E)$ presents a maximum near the point $E\rightarrow 0$. The pressure $P$ is inversely proportional to the energy $E$. The temperature presents a clear critical point $T_{0}=(4\pi d)^{-1} $. 
Superheating points for which $T\rightarrow \infty $,
do appear: at $E_{low}\approx \sqrt{T_{0}}$ {for low energies} and at $E_{high}\approx T_{0}$ {for high energies}. For high energies (large masses) the temperature of the
system is essentially the \textit{Hawking temperature} $T_{H}$ plus a (non logarithmic)
correction, while at low energies (small masses) the temperature is practically the string 
temperature $T_{0}$.  For $N>>1$ the specific heat behaves as $C_{v}\approx -1/(8 \pi bT^{2})$.
 
For $a\neq 5/2$, the particle density number N/V is a function of the energy density E/V.
The density of states $\Omega _{N}(E)$ presents a maximum near the point $E\approx
Nm_{0}$. The temperature presents the critical point $T_{0}= (4\pi d)^{-1}$ and  there are no superheating points. For $E\rightarrow nm_{0}$ (lowest energy of the system) as well as for $E\rightarrow \infty $, the temperature vanishes, while for  $E\rightarrow \ 0$, $T$ goes to the critical point $T\rightarrow \ T_{0}$. For $N>>1$, the temperature goes to the Hawking temperature $T_{H}$ plus a (logarithmic) correction. \
The pressure is negative with the form $P = - N T \ln N/({\frac{5}{2}-a})$. The specific heat $C_{v}$ is negative, for low energies it behaves as $C_{v} \approx {-1}/{8\pi b T_{0} ^{2}}\approx const.$, while for high energies $C_{v}$ vanishes, putting in evidence the asymptotic stability point of the system. \\

For strings, the parameter $a$ is related to the number of space-time dimensions $D$: $a=D$ (closed strings) or $ a=(D-1)/2$ (open strings). Thus, the relevant cases for the number of
dimensions of interest in string theory ($D=4, 10$ or $26$), are the cases $a\neq 5/2$, although it is useful for the sake of completeness, comparison and universality properties of the system to analyze the case $a=5/2$ too, as both cases appear together.

\begin{acknowledgements}

D.C.L. acknowledges the Observatoire de Paris, LERMA, for the kind hospitality extended to him and his Phd. supervisor E. A. Ivanov and  Prof. A. Dorokhov for their interest in this research.

\end{acknowledgements}

\subsection{References}

[1]  N.G. Sanchez, IJMP \textbf{A19}, 4173 (2004).\newline

[2]  M. Ramon Medrano and N. Sanchez, Phys. Rev. \textbf{D61}, 084030 (2000).\newline

[3]   A. Bouchareb, M.Ramon Medrano and N. Sanchez, IJMP \textbf {A22}, 1627 (2007).\newline

[4]  M. Ramon Medrano and N. Sanchez, Phys. Rev. \textbf{D60}, 125014 (1999).\newline
     A. Bouchareb, M. Ramon Medrano and N. Sanchez, IJMP \textbf{D6}, 1053 (2007).\newline

[5]  M. Ramon Medrano and N. Sanchez, MPLA\textbf{18}, 2537 (2003).\newline
     M. Ramon Medrano and N. Sanchez, MPLA\textbf{22}, 1133 (2007).\newline

[6]  J. D. Bekenstein, Phys. Rev. \textbf{D 7}, 2333 (1973).\newline

[7]  S. W. Hawking, Commun. Math. Phys. \textbf{43}, 199 (1975).\newline

[8]  R. Hagedorn, Nuovo Cimento \textbf{52 A}, 1336 (1967).\newline

[9] R. Hagedorn, Suppl. Nuovo Cimento \textbf{3}, 147 (1965).\newline

[10]  S. Frautschi, Phys. Rev. \textbf{D3}, 2821 (1971). \newline

[11]  S. Carlitz, Phys. Rev. \textbf{D5}, 3231 (1972). \newline

[12]  B. Harms and Y. Leblanc, Phys. Rev. \textbf{D46}, 2334 (1992).\newline

[13]  B. Harms and Y. Leblanc, Phys. Rev. \textbf{D47}, 2438 (1993).\newline

[14]  R. Casadio, B. Harms and Y. Leblanc, arXiv:gr-qc9706005.\newline

[15]  H. Umezawa, H. Matsumoto and M. Tachiki, \textit{Thermo Field Dynamics
      and Condensed States, }North Holland Publishing Co.,Amsterdam (1982).\newline

[16]  R. Dashen, S. Ma, H.J. Bernstein, Phys. Rev.\textbf{187}, 345 (1969).\newline

[17]  Yu. A. Brichkov and A. P. Prudnikov, \textit{Integral transform of
      General Functions, }Nauka, Moscow (1977).(In Russian).\newline

[18]  J. Scherk, Rev. Mod. Phys. \textbf{47}, 123 (1975).\newline

[19]  G. Veneziano, Phys. Rept. \textbf{9}, 199 (1974).\newline

[20]  L. Sertorio and M. Toller, N.C., \textbf{14 A}, 21 (1973).\newline

[21]  B.M. Barbashov, N.A. Chernikov, Zh. Eksp. Theor.Fiz. \textbf{50}, 1296 (1966).\newline
       Commun. Math. Phys. \textbf{5}, 313 (1966).

\end{document}